\documentclass[12pt]{article}
\usepackage{psfrag}
\newcommand{\be}{\begin{equation}}
\newcommand{\ee}{\end{equation}}
\newcommand{\beq}{\begin{eqnarray}}
\newcommand{\eeq}{\end{eqnarray}}
\newcommand{\la}{\langle}
\newcommand{\ra}{\rangle}
\begin{document}

\begin{center}
{\Large {\bf Critical Equalities for Potts Models } }
\end{center}

\vspace{1cm}

\begin{center}

{\sl P.Sawicki }
\vspace{0.5cm}

{\sl Institute of Physics, Jagiellonian University, \\
ul. Reymonta 4, PL-30-059, Krak\'ow }

\end{center}

\vspace{2cm}
\begin{center}
{\bf Abstract }
\end{center}

We apply  a simple analytical criterion for locating critical temperatures
to Potts models on  square and triangular lattices. In the self-dual
case, i.\ e. on the square lattice we reproduce known exact values of
the critical temperature  and derive the internal energy of the model
at the critical point. For the Potts model on the triangular lattice 
we obtain  very good numerical estimate of the critical temperature 
and also of the internal energy at the critical point.

\vspace{3cm}
\noindent
Keywords: Potts model, Phase transition, Transfer matrix \\
PACS numbers: 75.10.Hk \\

\newpage

\section{Introduction}
Critical temperature is  known only for  very few spin models 
like the two-dimen\-sio\-nal Ising model, which is exactly 
soluble  in $2d$, and Potts models where the critical 
temperature can be found by the duality relations [1-3]. 
In principle a position of the critical point can be obtained
using one of many existing approximate techniques ranging from  
high and low temperature expansions to renormalization group methods.   
In this paper we employ  a simple method  based on the moments of the
transfer matrix \cite{wosiek}.  Let ${\cal T}$ be the transfer matrix
for a $d$-dimensional spin system with linear size $L$ and let  
$\beta=1/k_B T$. One defines a characteristic function of rank $n$
\be
\rho_{n} (\beta )= \lim_{L \rightarrow \infty } {\left(  \frac
        {{(Tr {\cal T})}^{n} }
        {Tr {\cal T}^{n} } \right)}^{\frac{1}{L^{d-1} }}. \label{one}
\ee
As was conjectured in \cite{wosiek} the maximum of $\rho_n$ 
should occur at the critical point 
\be
\beta_{max} = \beta_{crit}.       \label{max}
\ee

The existing results suggest that the validity of the method is related
to the dual properties of the system. It was checked directly for
the Ising and Potts models with low $q$ values on the square lattice
that the criterion reproduces exact values of the critical temperature. 
It can be made exact also for the isotropic Ashkin-Teller model, for which 
the duality transformation involves three couplings. However,
in this case the definition of the characteristic function should be modified 
[5,6]. On the other hand, for the two-layer Ising model, which is not
self-dual, the criterion still gives surprisingly good numerical
estimate of the critical temperature \cite{burda}. Therefore it is 
very tempting  to perform further investigations of the method to test
its applicability in different situations. This is a part of our programme
whose final goal is an exact criterion for not self-dual systems.

In this paper we test the proposal (Eq. (\ref{max})) for the 
two-dimensional Potts model. Contrary to the previous mainly numerical
work \cite{wosiek} our discussion is based on the explicit analytical
expression for the function $\rho_2$ for arbitrary $q$. We also check
some consequences arising from the criterion in the limiting case 
$n \rightarrow \infty$.  These are interesting equalities relating
internal energies of one and two dimensional systems. The calculation
of $\rho_2$ is given in Section 2 and the  equalities mentioned above
are derived in Section 3. Appendix contains some technical details.

\section{The $\rho_2$ function }

The partition function of $n$ coupled chains is related to 
the matrix $T_n$, which propagates horizontally $n$ spins
\be
Z_n=Tr {\cal T}^n= Tr (T_{n})^{L} ,
\ee 
where $L$ is the linear extension of the system as in the formula
(\ref{one}). The construction of the matrix $T_n$ is shown, for $n=2$, in
Fig.\ 1 where periodic boundary conditions are assumed in vertical
direction. Since in the thermodynamic limit ($L \rightarrow \infty$)
only the largest eigenvalue, $t_{n,max}$, of the matrix $T_n$
dominates, we obtain for $\rho_n$ the expression
\be
\rho_n (\beta) = \frac{ t_{1,max}^{n} }{t_{n,max} },
\ee
which is very convenient for analytical calculations.

For the Potts model the matrices $T_1$ and $T_2$ have a simple form
and it is relatively easy to find analytical expressions for 
the largest eigenvalues. This is done in the Appendix where the 
formulas for the $\rho_2$ function are given for square 
and  triangular lattices. The functions $\rho_2(x)$, where $x=e^\beta$,
for the Potts model on the square lattice  are shown in Fig.\ 2.
For arbitrary $q$ they have the same characteristic shape with the maximum,
whose height and position depend on $q$. The position of the maximum
is located at $x= 1 + \sqrt{q}$ which agrees with the standard values
obtained by duality of the six vertex model [1,3]. Indeed at this point 
the Potts model undergoes the second order transition for $q \leq 4$
and first order for $q>4$. 

The triangular lattice is not self-dual but it is dual to
the honeycomb lattice. The exact critical temperature still can be
found from some properties of ice-type models and it is given 
as a root of the cubic equation [1,3]
\be
x^3 - 3 x -q + 2 = 0 .
\ee
The results for this case are summarized in Tab.\ 1, where we compare 
the critical values and the position of the maximum of $\rho_2$. The 
criterion still gives a good estimation of the critical temperature. 
In particular for $q=2$ it is exact, which is related to the existence 
of the star-triangle relation which provides a certain kind of 
duality transformation.

\section{Critical equalities}
By taking the  logarithm of (Eq.\ 1) and by differentiating it with respect
to $x$ one obtains the condition for the internal energies 
\be
u_1 (x_{max})= u_n (x_{max}),
\ee
where $u_1$ ($u_n$) denotes the density of the internal energy for a spin
chain ($n$ coupled spin chains). In the limit $(n \rightarrow \infty)$ we 
conclude that at the critical point the energy density of the $2d$ system 
equals to the energy density of a single chain. This allows us to use
standard expressions for the internal energy in Potts models  
to check validity of our criterion also in this limiting situation.

Let us first consider the square lattice. The internal energy
of a chain per one spin is:
\be
u_{1}= \frac{2x +q -1}{x+q-1}.
\ee
Evaluating this at the critical point $x_{crit}= 1 + \sqrt{q}$ we obtain,
in agreement with known results [1,3], namely that the energy of the
$2d$ system is:
\be
u_{crit} = 1 + \frac{1}{\sqrt{q} }.
\ee
For $q > 4$ when the system undergoes the first order phase
transition the internal energy at the critical point is discontinuous.
To fix this ambiguity one usualy defines the energy as a mean 
of the high and low temperature phase, and so should be interpreted
our results. 
 
In the last three rows of Tab.\ 1 we compare the predictions for
numerical values of internal energies for the Potts system on the
triangular lattice. $E(T_c)$ and $L$ are  results for
the internal energy and latent heat respectively taken from
Ref. \cite{wu}. $E(T_{max})$ is the energy of a single chain evaluated
at the maximum point. The method still gives  good
numerical estimation of the internal energy. For the first 
order phase transition it fits nicely between energy of the high and 
low temperature phase.

The critical equalities are simple mathematical consequences
of the maximum rule but their physical origin is unclear. 
One can say that at the critical point when the correlation length
$\xi$ diverges the interaction of a single chain with the rest of the
$2d$ system can be simplified to the interaction  of spins in a single
chain with periodic boundary conditions. This would also mean that
characteristic function should contain information about correlations
in the system. Thus it should be possible to obtain other
characteristics of the  transition, for example to  distinguish the
order of  the transition and to find the critical exponents.  One can
may be improve the method by combining it with the
renormalization group techniques. It would also be tempting to 
look for signals of the order of transition containing in the
function $\rho_n$. For the Potts model shape of $\rho_2$ exhibits 
rather slow crossover as a function of $q$ and for the moment it is
not clear weather one can see any particular change at $q=5$.

\vspace{1cm}
\centerline{\Large Acknowledgements}
I would like to thank the Foundation for Polish Science for a fellowship.
I am very grateful to J. Wosiek for fruitful discussions and Z. Burda for
reading the manuscript. Work supported in part by the KBN grants
no: 2P03B19609 and 2P03B04412.

\section{Appendix}
In this Appendix we derive analytical formulas for the
$\rho_{2}$ function of the $2d$ Potts model.
The energy of the $q$ state Potts model is given by
\be
E(s) = - \sum_{ \la i,j \ra} \delta_{s_i , s_j} ,
\ee
where the sum runs over nearest neighbours sites labeled by $i,j$ and
the spin variables $s$ are assumed to take $q$ distinct values. The transfer
matrices can be specified by their elements between spin states. For
the square lattice they are:
\beq
\la s | T_1 | s' \ra & = & \exp \beta ( \delta_{s,s'} +1 ), \nonumber \\
\la s_1 s_2 | T_2 | s_{1}^{'} s_{2}{'} \ra & = & 
\exp \beta (\delta_{s_1, s_2}
+ \delta_{s_1, s_{1}^{'} } + \delta_{ s_2 s_{2}^{'} }
+ \delta_{ s_{1}^{'} s_{2}^{'} } ) .
\eeq

It can be readily verified that  $t_{1 max}=x(x+q-1)$. The largest
eigenvalue of the matrix $T_2$ can be
obtained noting the special structure of the Potts interaction. To make
it more evident we introduce an index  $n=0,\ldots,q^2-1$, which 
enumerates successive rows (columns) of the matrix. $n$ can be represented 
as a two digit number in base $q$.  With each digit we
associate a spin variable. Let us now construct the
equation for the eigenvalues. The $0$-th, $(q+1)$-th,\ldots,$(q^{2}-1)$-th
row of the resulting determinant we multiply by some constant, say $A$.
At the end we add all the rows to the last one.  It is easy to notice that
we generically obtain two expressions in the last row: in  the $0$-th,
$(q+1)$-th,\ldots,$(q^{2}-1)$-th column $w_1$ and in the other columns $w_2$.
This is because the Potts interaction depends on  whether
two spins are equal or different. Detailed form of $w_1$ and $w_2$
can be obtained from some combinatorical considerations
\beq
w_{1} = A(x^{4}-\lambda + (q-1)x^{2}) + (q-2) (q-1)x +2(q-1) x^{2} ,\\
w_{2} = A(2 x^{2} +(q-2) x) +x^{2}-\lambda +2(q-2)x +q^2-3q+3.
\eeq

If $w_{1}=w_{2}=0$ one obtains a quadratic equation for a certain
eigenvalue. We choose the larger root of this equation
\beq
\lambda_{1}(x)= \frac{1}{2} (Q_{2}(x) +\sqrt{Q_{3}(x)} ),
\eeq
where $Q_{2}$ and $Q_{3}$ are polynomials given below. 
The other eigenvalues $\lambda_i (x) \; i=2,\ldots,q^2$
can be considered as continuous functions of $x$. 
Moreover $\lambda_1(1)=q^2$, $\lambda_i (1) =0$. By the Perron's theorem 
\cite{perron}, $\lambda_1 (x) \neq \lambda_i (x)$ since the elements
of $T_2$ are positive. Thus $\lambda_1 (x)$ must remain the largest
eigenvalue in the whole interval $(1, \infty)$.

Collecting our results together we finally obtain:
\beq
 \rho_2 (\beta )=\frac{Q_{1} (x)} {Q_{2}(x) + \sqrt{Q_{3}(x)} },
  \label{forma}
\eeq
where
\beq
Q_{1}(x)  & = & 2 x^{2} (x+q-1)^{2} \label{sq1}, \\
Q_{2}(x)  & = & x^{4} + q x^{2} +(2q-4)x +q^{2}-3q +3 , \label{sq2} \\
Q_{3}(x)  & = & x^{8} +(2q-4) x^{6} + (8-4q) x^{5} +(-q^{2} +18q -18)
            x^{4} \nonumber \\
      &   & +(12 q^{2}-32q+16) x^{3}  +(2 q^{3} -6 q^{2} -2 q +12)
            x^{2} \nonumber \\
      &   &  +(4q^{3}-20 q^{2} +36 q -24) x +(q^{2}-3 q+3)^{2} \label{sq3}.
\eeq

The method described above applies also to the transfer matrices of the
Potts model on the triangular lattice
\beq
\la s |T_1 s \ra & = & \exp{\beta(2 \delta_{s s'}+1) }   \nonumber \\
\la s_1 s_2 | T_2 | s_{1}^{'} s_{2}{'} \ra & = & \exp 
\beta (\delta_{s_1, s_2}
+ \delta_{s_1, s_{1}^{'} } + \delta_{ s_2 s_{2}^{'} }
+ \delta_{ s_{1}^{'} s_{2}^{'} } +\delta_{s_{1} s_{2}^{'} } +
\delta_{s_2 s_{1}^{'} } ) . 
\eeq
The final result has the same form as (\ref{forma}) with
\beq
Q_{1} (x) & = & 2 x^2 (x^2 +q-1)^2 \label{tr1}, \\
Q_{2} (x) & = & x^6 + x^2 (q+1) +x ( 4 q -8) +q^2 -5 q +6, \label{tr2} \\
Q_{3} (x) &= & x^{12} + ( 2q-6 ) x^8 + (- 8q + 16) x^7 + \nonumber \\
          &  & (-2 q^2 + 26q - 28) x^6 + (17 q^2 - 54q + 41) x^4 \nonumber \\
          &  &  + (-8 q^2 + 40q - 48 ) x^3
               +(2 q^3 +12 q^2 - 74q + 84 ) x^2 \nonumber \\
          &  &  +( 8 q^3 - 56q^2 + 128 q -96) x \nonumber \\
          &  & q^4 -10 q^3 +37 q^2 -60 q +36.  \label{tr3}
\eeq

\newpage
\begin{table}
\begin{tabular}{|l|l|l|l|l|l|l|l|l|}
\hline
q & 2 & 3 & 4 & 5 & 6 & 7 & 8 & ...100 \\
\hline
$e^{\beta_{cr}} $ & $\sqrt{3}$ & 1.8794 & 2 & 2.1038 & 2.1958 & 2.2790
& 2.3553 & 4.8272 \\
\hline
$e^{\beta_{max}} $ & $\sqrt{3}$ & 1.8757 & 1.9930 & 2.0938 & 2.1893 & 2.2635
 & 2.3373 & 4.7266 \\
\hline
$E(T_{c})$ & 0.8333 & 0.7603 & 0.7131 & 0.6881 & 0.6669 & 0.6506 &0.6377 &
$\star$ \\
\hline
$L(q)$ & 0 & 0 & 0 & 0.0310 & 0.1172 & 0.2042 & 0.2795 & $\star$ \\
\hline
$E(T_{max})$ &0.8333 & 0.7584 & 0.7026 & 0.6819 & 0.6596 & 0.6404 & 0.6256 &
0.4561 \\
\hline
\end{tabular}
\caption{ Comparison of  the results taken from Ref. [3] and the
values obtained from the maximum criterion for the $2d$ Potts model on
the triangular lattice. }

\end{table}

\begin{figure}

\includegraphics{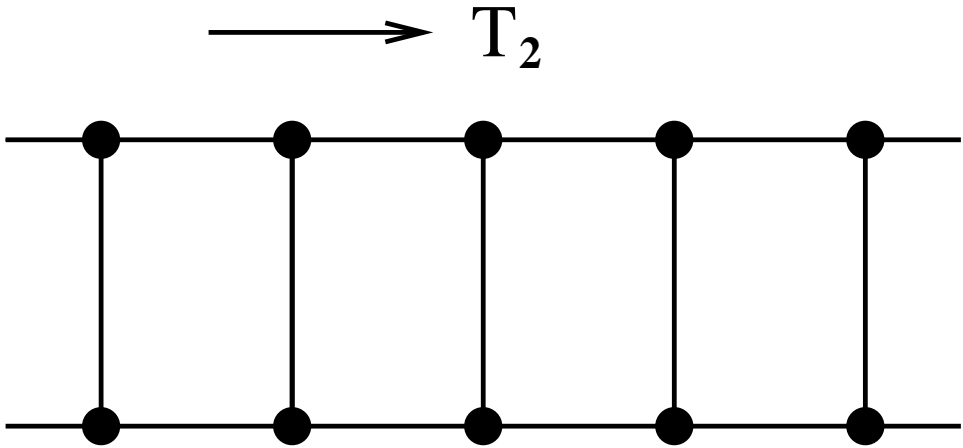}

\caption{Construction of the transfer matrix $T_2$ for the
two-dimensional Potts model on the square lattice. Periodic boundary
conditions are assumed in vertical direction. }
\end{figure}

\begin{figure}
\psfrag{yy}{ {\Large $\rho_2(x) $ } }
\includegraphics{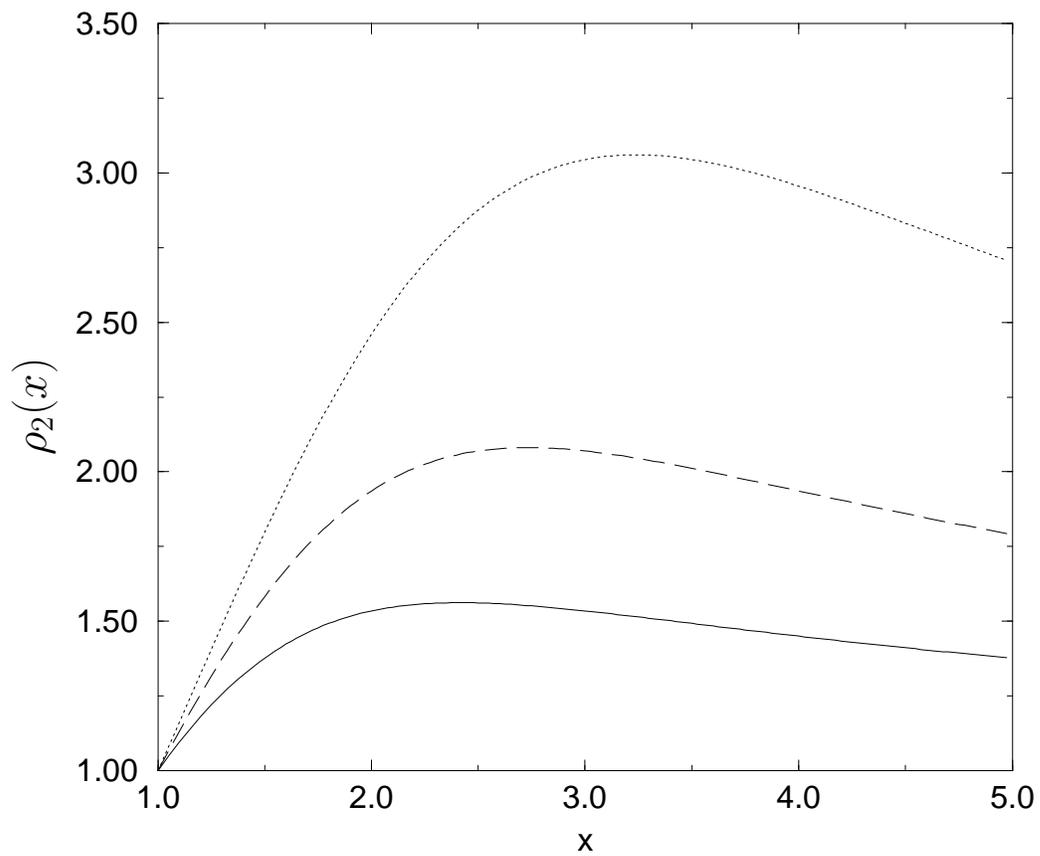}
\caption{ Functions $\rho_2$ for the Potts model
on the square lattice. Three lines are shown for $q=2$ (solid)
, $q=3$ (dashed) and $q=5$ (dotted). } 
\end{figure}

\end{document}